\documentclass[10pt,twocolumn]{elsart}
\usepackage{natbib}
\usepackage{graphicx}
\begin{document}
\begin{frontmatter}
\title{Digital watermarking in the singular vector domain}
\author{Rashmi Agarwal},
\author{M.S. Santhanam}\footnote{Corresponding author.}
\ead{santh@prl.res.in}
\address{Physical Research Laboratory, Navrangpura, Ahmedabad-380 009, India}
\begin{abstract}

Many current watermarking algorithms insert data in the spatial or
transform domains like the discrete cosine, the discrete Fourier,
and the discrete wavelet transforms. In this paper, we present a
data-hiding algorithm that exploits the singular value decomposition
(SVD) representation of the data. We compute the SVD of the host
image and the watermark and embed the watermark in the singular
vectors of the host image. The proposed method leads to an
imperceptible scheme for digital images, both in grey scale and
color and is quite robust against attacks like noise and JPEG
compression.
\end{abstract}
\end{frontmatter}

\section {Introduction}

In the past one decade, there has been a phenomenal increase in the
use and circulation of information in digital multimedia formats for
various purposes. Today, many paintings, photographs, newspapers,
books, music etc., are available over the internet in one or the
other multimedia format. The increasing necessity to protect the
intellectual property rights of such digital content has led to
considerable research in that direction. Digital watermarking is one
such approach. Watermarking is the process of embedding data into a
multimedia element such as an image, audio or video [Langelaar et
al. (2000)]. This embedded data can later be extracted from, or
detected in, the multimedia for several purposes including copyright
protection [Craver et al. (1998), Zeng and Liu (1999)], trade marks,
access control or even to simply pass a secret piece of information
hidden in an innocuous digital image [Zeng (1998), Petitcolas et al.
(1999)]. A watermarking algorithm consists of the watermark
structure, an embedding algorithm and an extraction and detection
algorithm.

In this paper, we will be concerned with invisible watermarking of
the digital images. In such images, the basic idea is to embed the
watermark image in a given host image such that the resultant
(watermarked) image carries the watermark either in visible or
invisible mode. It is important that the watermarked image should
suffer least corruption due to the watermarking procedure. The broad
strategy, then, is to embed the watermark, preferably, in the least
significant part of the transformed domain of the image. Techniques
that employ this strategy are the discrete cosine transformation
(DCT) [Bors and Pitas (1996), Dickinson (1997), Piva et al. (1997),
Piva et al. (1998)] and the wavelet transform [Kundur and Hatzinakos
(1997), Kundur and Hatzinakos (1998)] that also happen to be most
popular [Katzenbeisser and Petitcolas (2000)]. The DCT allows an
image to be broken up into different frequency bands, making it much
easier to embed watermarking information into the middle frequency
bands of an image. These bands are chosen such that they minimize
the changes to the important parts of the host image and remain
unaffected by various image transformations [Langelaar et al.
(2000)]. The wavelet transform approach embodies a similar
philosophy. The transformed domain method, such as this, works well
also because the embedded data is located in that part of the
frequency that is least sensitive for the human visual system hence,
the watermark is imperceptible [Lacy et al. (1998), Prandoni and
Vetterli (1998), Ohbuchi et al. (2002)].

However, notice that transform methods such as the DCT or wavelets
attempt to decompose the image in terms of a standard basis set.
This need not necessarily be the optimal representation for a given
image. Singular value decomposition offers a method by which the
transformed domain consists of basis states that is optimal in some
sense; i.e., a tailor-made basis for a given image. Many of the
earlier works using SVD for watermarking have attempted to
manipulate the singular values of the host and the watermark images
one way or the other [Liu and Tan (2002), Shieh et al. (2005)].
Another approach is to perform an SVD on various blocks of the image
and then add the scaled singular values of the watermark image to
that of the host image [Gorodetski et al. (2001)]. For a survey of
existing watermarking methods using SVD, we refer the reader to
Ganic et al. (2003). In the scheme proposed by Chang et al. (2005),
both the singular values and singular vectors are explored for
embedding the watermark. A watermarking scheme using minimax
eigenvalue decomposition has been proposed by Davidson and Allen
(1998). In our algorithm, we assume a situation in which those who
have opted to watermark their digital images have access to their
original images (without the watermark) and the watermark image for
verification purposes later. This does not restrict the
applicability of our technique as we argue below. In the proposed
method we work at the level of singular vectors and embed the
watermark in the singular vectors of the host image. The singular
vectors of an image of size ($m>n$) $m \times n$ has $n^2$
parameters and contains detailed, graded information about the image
as opposed to just $n$ singular values. Hence, in the domain of
singular vectors, we have more latitude to embed the watermark. We
argue that this also leads to a level of digital security in
watermark embedding. We also subject our algorithm to tests against
additive noise, cropping and JPEG compression.

In the next section, we recall the singular value decomposition and then
describe our algorithm. Then, we present our numerical simulations including
the robustness to several possible attacks.

\section{SVD-Based Watermarking}

Singular value decomposition is a popular technique in linear
algebra and it has applications in matrix inversion, obtaining low
dimensional representation for high dimensional data, for data
compression and even data denoising etc., [Golub and Reinsch (1970),
Andrews and Patterson (1976), Leon (1994)]. If ${\mathbf Z}$ is any
$m \times n$ matrix, it is possible to find a decomposition of the
form
\begin{equation} \label{svd}
{\mathbf Z} = \mathbf{U~D~V}^{\mbox{T}},
\end{equation}
where ${\mathbf U}$ and ${\mathbf V}$ are orthogonal matrices of
order $m \times n$ and $n \times n$ respectively. Proof of
Eq.(\ref{svd}) can be found in many standard linear algebra
literature [ Strang (1993)]. The diagonal matrix ${\mathbf D}$ of
order $n \times n$ has elements $d_{ii}, (i=1,2,..n)$, which are
positive definite and are called the singular values of ${\mathbf
D}$. We do not consider here the conditions under which a
decomposition of type in Eq. (\ref{svd}) will fail for image
matrices and hence we assume that, for all practical purposes, the
image matrix ${\mathbf Z}$ can be decomposed in the form given in
Eq. (\ref{svd}). Thus the SVD can be applied directly to digital
images represented as matrix arrays.

\subsection{Algorithm for embedding watermarking in gray scale image}

In this section we present our algorithm to embed a gray scale image
into another gray scale image of the same size using SVD. Let the
matrix $\mathbf{Z}$ represent the host image which needs to be
watermarked. Let $\mathbf{W}$ represent the matrix of the image to
be embedded. As a first step, we compute the SVD of both
$\mathbf{Z}$ and $\mathbf{W}$.
 \begin{equation}
\mathbf{Z}=\mathbf{U_z~D_z~V_z^{\mbox{T}}} = \mathbf{A_z ~
V_z^{\mbox{T}}} \label{eq2}
 \end{equation}
For the watermark image
\begin{equation}
\mathbf{W}=\mathbf{U_w~D_w~V_w^{\mbox{T}}} = \mathbf{A_w ~
V_w^{\mbox{T}}} \label{eq3}
\end{equation}
where $\mathbf{A_{z/w} = U_{z/w}~D_{z/w}}$ are also called the
principal components in the language of principal component
analysis.

Now, we add the scaled eigenvector $\mathbf{V_w}$ of watermark to that
of the original image,
\begin{equation} \label{orimage}
\mathbf{V} = \mathbf{V_z} + \lambda \mathbf{V_w}
\end{equation}
where $\lambda$ is the scaling factor. Typically, $0 \le  \lambda
\le 1$, so that the intensity of the watermark $\mathbf{W}$ is less
compared to the original image $\mathbf{Z}$. Note that, within the
framework of SVD, $\mathbf{V_w ~V_w^{\mbox{T}} = I}$, where
$\mathbf{I}$ is the identity matrix. Similar relation holds good for
$\mathbf{V_z}$ too. As $\lambda \to 0$, the approximation that
$\mathbf{V}$ is a orthogonal matrix, i.e, $\mathbf{V~V^{\mbox{T}}
\approx I}$ gets better. This property is important in the next step
for constructing the watermarked image. We get the watermarked image
as,
\begin{equation}
\mathbf{Z_c}= \mathbf{A_z ~ V^{\mbox{T}}}
\label{img_zc}
\end{equation}
Thus, equations (\ref{eq2}-\ref{img_zc}) constitute the algorithm
for watermarking using SVD in the eigenvector domain.

\subsection{Algorithm for extracting watermarks}

Given the watermarked image  $\mathbf{Z_c}$, we can extract,
possibly a corrupted watermark, if we have access to the matrices
$\mathbf{A_z, A_w, V_z}$ and the value of $\lambda$. That is, we
assume that whoever wants to extract the watermark should have
access to the original image as well as the embedded watermark
image. We emphasize that this is not a restrictive assumption. In
most cases, involving copyrights and trademarks embedded in digital
images, the person or organization that embedded the watermark in
their proprietary digital image will have access to both the clean
original image and the watermark image.

Extraction algorithm is a straightforward reversal of the embedding
algorithm given by equations (\ref{eq2}-\ref{img_zc}). Starting from
Eq.(\ref{img_zc}), we multiply both sides of Eq. (\ref{img_zc}) by
$\mathbf{A_z^{\mbox{-1}}}$ and substitute for
$\mathbf{V^{\mbox{T}}}$ from Eq. (\ref{orimage}). It is
straightforward to obtain an expression for
$\mathbf{V_w}^{\mbox{T}}$ as
\begin{equation} \label{vw}
\mathbf{V_w^{\mbox{T}}} = \frac{ \mathbf{A_z^{\mbox{-1}} Z_c} -
\mathbf{V_z^{\mbox{T}}} }{\lambda}.
\end{equation}
Finally, using Eq. (\ref{eq3}), the watermark image can be
constructed as,
 \begin{equation}
\mathbf {\widetilde{W}} = \mathbf{A_w} \mathbf{V_w^{\mbox{T}}}.
\label{eq7}
 \end{equation}
Eq. (\ref{eq7}), along with Eq. (\ref{vw}) constitutes the watermark
extraction algorithm.

In general, the matrix $\mathbf{A_z}$ is not a square matrix and by
Eq. (\ref{vw}) we are required to take its inverse. This computation
is simple because from Eq. (\ref{eq2}), we have, $\mathbf{A_z^{-1}}
= \mathbf{D_z^{-1}~U_z^T}$, where inverse of a diagonal matrix and a
transpose need to be computed. At this point, we also stress that if
$d_{ii} > 0$, then $\mathbf{A_z^{-1}}$ exists, even if
$\mathbf{A_z}$ is not a square matrix.

Firstly, note that even though we simulate with square images, the
SVD based method presented above can handle rectangular images as
well, without any changes to the algorithm. One important extension
of this algorithm is to address the case of color images. We
consider the case of RGB coded color images. If a color image is
specified in RGB format then it comprises of three component
matrices superimposed together, one each for red, green and blue. It
can be written as, $\mathbf{Z} = \mathbf{Z_r + Z_g + Z_b}$ and
similarly for the watermark image $\mathbf{W}$. We can directly
apply this algorithm to each of the respective component images,
$\{\mathbf{Z_r,W_r}\}$,  $\{\mathbf{Z_g,W_g}\}$ and
$\{\mathbf{Z_b,W_b}\}$.

Secondly, the algorithm provides certain implicit security features
as well. The extraction algorithm given by Eq. (\ref{vw}) requires
the knowledge of $\mathbf{A_z,V_z,A_w}$. Notice that by performing a
SVD on the watermarked image $\mathbf{Z_c}$, it might be possible to
obtain approximate estimates of $\mathbf{A_z}$ and $\mathbf{V_z}$.
However, unless one knows $\mathbf{A_w}$ and $\lambda$, it is not
possible to extract the watermark. This means that if $\mathbf{Z_c}$
is an $m \times n$ array, then to hack the embedded watermark, one
has to estimate $mn+1$ independent parameters. This becomes further
complicated due to the fact that the watermark is embedded uniformly
in the entire host image as given in Eq.(\ref{orimage}). This is in
contrast to the methods [Chandra (2002)] where the information about
the watermark is embedded only in $n$ values of the original image.

\section{Numerical simulations for gray scale image}

\begin{figure}[h]
\begin{center}
\begin{tabular}{cc}
\begin{minipage}{1.2in}
\centering
\includegraphics[width=1.0in,height=1.0in]{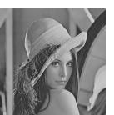}
\caption{Original image $128 \times 128$}\label{e1}
\end{minipage}
&
\begin{minipage}{1.2in}
\centering
\includegraphics[width=1.0in,height=1.0in]{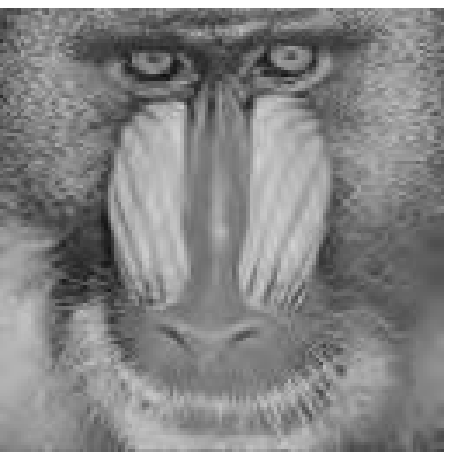}
\caption{Watermark image $128 \times 128$}\label{e2}
\end{minipage}
\end{tabular}
\end{center}
\end{figure}

\begin{figure}[h]
\begin{center}
\begin{tabular}{cc}
\begin{minipage}{1.2in}
\centering
\includegraphics[width=1.0in,height=1.0in]{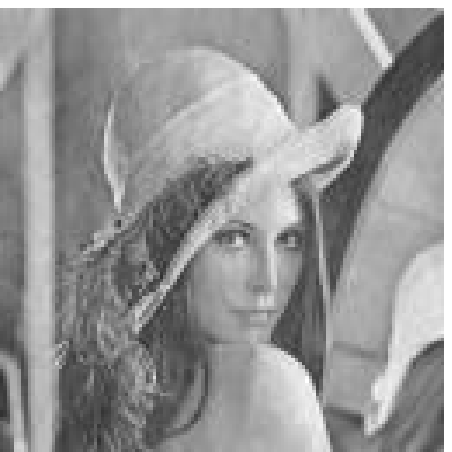}
\caption{Watermarked image} \label{e3}
\end{minipage}
&
\begin{minipage}{1.2in}
\centering
\includegraphics[width=1.0in,height=1.0in]{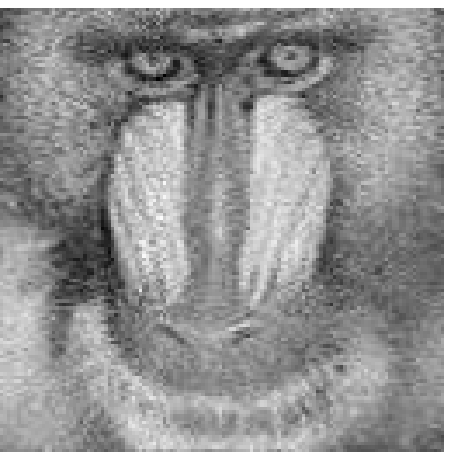}
\caption{Recovered watermark image} \label{e4}
\end{minipage}
\end{tabular}
\end{center}
\end{figure}

\begin{figure}[h]
\begin{center}
\includegraphics[width=1.0in,height=1.0in]{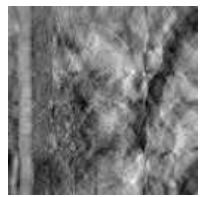}
\end{center}
\caption{Image difference $\mathbf{\Delta_z}$ between Fig.(1) and
Fig.(3) for $\lambda=0.2$.} \label{e5}
\end{figure}

\begin{figure}[h]
\begin{center}
\includegraphics[width=2in,height=2in]{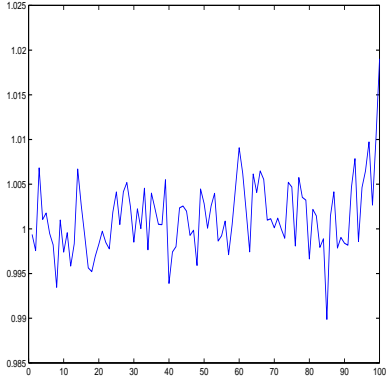}
\end{center}
\caption{Linear graph of diagonal element of the resultant vector
 ($VV^T$)} \label{e6}
\end{figure}

In Fig. (\ref{e1}) we show the original Lena image, the watermark in
Fig. (\ref{e2}), the watermarked image in Fig. (\ref{e3}) and the
extracted watermark in Fig. (\ref{e4}). All the images are arrays of
size $128 \times 128$ and $\lambda=0.2$. In Fig. (\ref{e5}), we show
the absolute difference between the original image and the
watermarked image, i.e, $\mathbf{\Delta_z} = |\mathbf{Z_c - Z}|$.
The embedded watermark is completely invisible and we only see a
texture of the original image. Even for larger values of $\lambda$,
we obtain a similar result. This means that even at 50\% strength of
watermark image in the host image, this method works well. As
$\lambda \to 0$, the approximation that $\mathbf{V}$ is a orthogonal
matrix, i.e, $\mathbf{V~V^{\mbox{T}} \approx I}$ gets better, is
shown by the linear graph of diagonal elements in Fig.(\ref{e6}).

\begin{figure}
\centering
\includegraphics[width=2.5in]{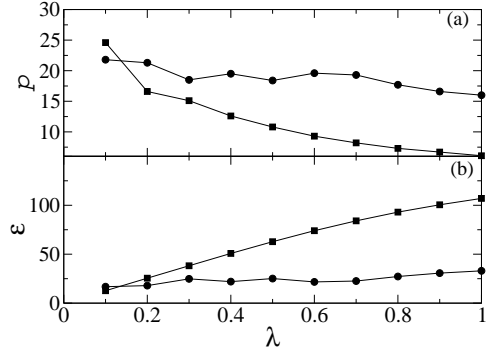}
\caption{Error between the original Lena image and the watermarked
image (squares). Error between the watermark and extracted watermark
(circles). (a) PSNR ($p$) and (b) RMSE ($\epsilon$) as a function of
$\lambda$.} \label{e7}
\end{figure}

In order to obtain a global picture of how this scheme performs as
the scaling factor $\lambda$ is increased, we compute the
root-mean-square error (RMSE) and signal-to-noise ratio (PSNR) for
the image difference matrix $\mathbf{\Delta_z}$ for each value of
$\lambda$. If $\delta_{x,y}$ represent the elements of
$\mathbf{\Delta_z}$, then the RMSE is defined as,
\begin{equation} \label{rmse}
\varepsilon=\sqrt{\frac{1}{mn}\sum_x^m \sum_y^n \delta_{x,y}^2}.
\end{equation}
For the case of square image arrays considered in our simulations,
$m=n$. We also define the Peak Signal-to-Noise Ratio (PSNR) as, $p =
10 \log_{10}(\mbox{max}_{x,y}/\varepsilon)$. In Fig. (\ref{e7}), we
show the $p$ and $\varepsilon$ as a function of $\lambda$. In this
figure, the scaling factor $\lambda$ increases, there is an
approximately linear increase in error as is to be expected. Beyond
about $\lambda=1.2$, the watermarked image becomes highly noisy.
RMSE can be interpreted as the average error per pixel and it is
seen that the error is only a small fraction of the pixel values.

\section{Robustness of the algorithm}

An important property of the watermarking algorithms is that they
should be robust against several kinds of attacks. In this section,
we show the results for (i) robustness against additive noise (ii)
cropping (iii) JPEG compression.

\subsection{Additive Noise}

The noisy image $\mathbf{Z^{'}}$ can be represented as,
\begin{equation}
\mathbf{Z^{'}=Z} + \mathbf{G}
\end{equation}
where $\mathbf{G}$ is a random matrix of same order as $\mathbf{Z}$
with entries drawn from a standard Gaussian distribution
$N(0,\sigma)$, where $\mathbf{\sigma}$ is the variance.
\begin{figure}[h]
\begin{center}
\includegraphics[width=1.0in,height=1.0in]{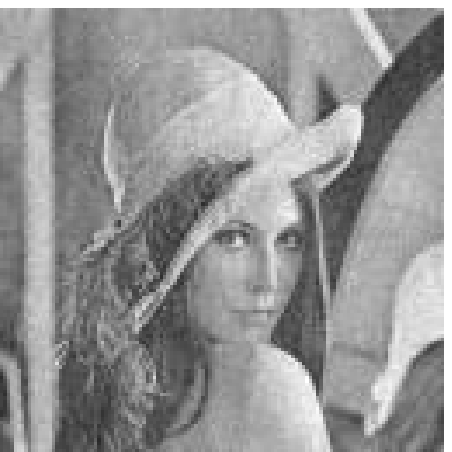}
\hskip 5mm
\includegraphics[width=1.0in,height=1.0in]{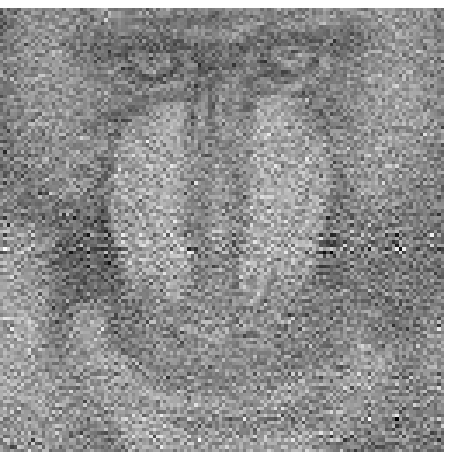}
\caption{Noisy watermarked image with $\lambda=0.2$ (left) and
recovered watermark (right). The PSNR error between watermark and
the extracted watermark is $p=16.3$ for $\sigma=0.04$.} \label{e8}
\end{center}
\end{figure}

\begin{figure}
\centering
\includegraphics[width=2.5in]{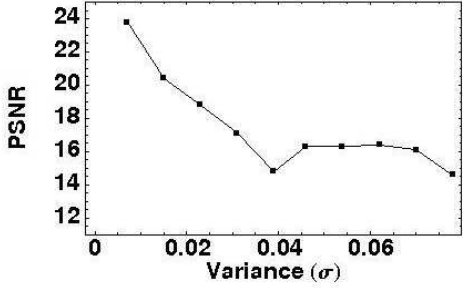}
\caption{ PSNR between the watermark and extracted watermark for
different variance $(\sigma)$.} \label{e9}
\end{figure}

Fig. (\ref{e8}) shows the result of extracting watermark from the
noisy Lena image. Our simulations indicate that the image difference
between watermark image and the one extracted from noisy image
correlate well for a large range of $\lambda$. Fig. (\ref{e9}) shows
PSNR between the watermark and extracted watermark for different
variance $(\sigma)$. It clearly indicates that as the variance of
the Gaussian noise is increased the corresponding PSNR value
decreases thereby leading to a deterioration in the quality of the
extracted image.

\subsection{Cropping}

\begin{figure}[h]
\begin{center}
\includegraphics[width=1.0in,height=1.0in]{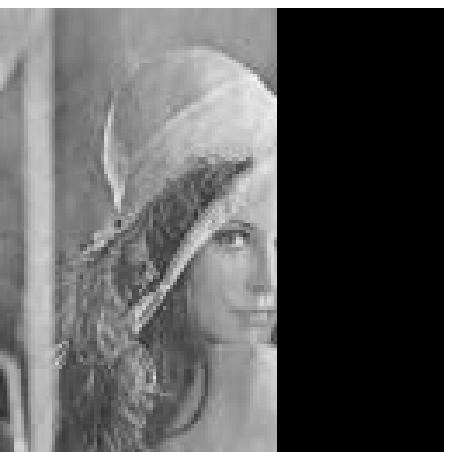}
\hskip 5mm
\includegraphics[width=1.0in,height=1.0in]{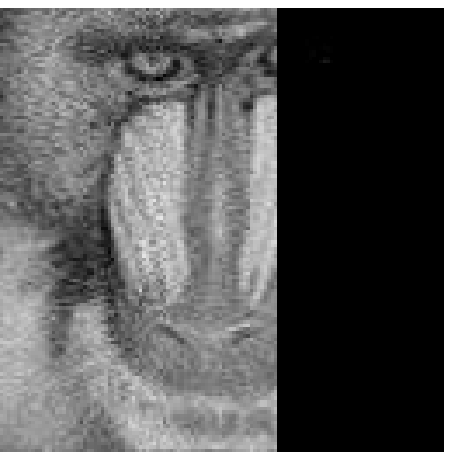}
\caption{Cropped Lena image (left) and the recovered watermark
(right).} \label{e10}
\end{center}
\end{figure}

Cropping is the process of removing certain parts of an image. Here,
we subject the original Lena image to cropping and the cropped image
is displayed in Fig. (\ref{e10}). The extracted watermark is also
displayed alongside. We recall that in this method the information
about the watermark is stored in every part of the original image.
When the watermarked image is cropped and subjected to an extraction
procedure, then the watermark from the uncropped part survives. By
design, there is some loss of information. However, the surviving
part of the extracted watermark is very similar to the original
watermark.

\subsection{JPEG Compression}

\begin{figure}
\centering
\includegraphics[width=2.5in]{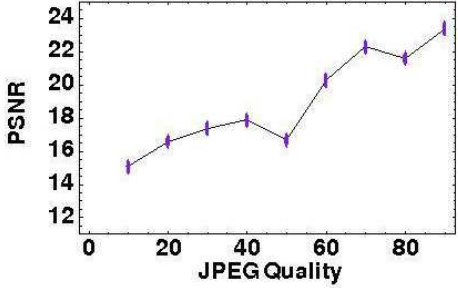}
\caption{ PSNR between the watermark and extracted watermark after
JPEG compression.} \label{e11}
\end{figure}

\begin{figure}
\begin{center}
\begin{tabular}{cc}
\begin{minipage}{1.2in}
\centering
\includegraphics[width=1.0in,height=1.0in]{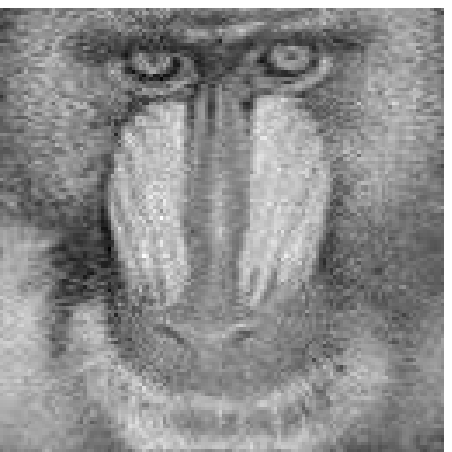}
\caption{JPEG Quality 90 ($p=23.40$)} \label{e12}
\end{minipage}
&
\begin{minipage}{1.2in}
\centering
\includegraphics[width=1.0in,height=1.0in]{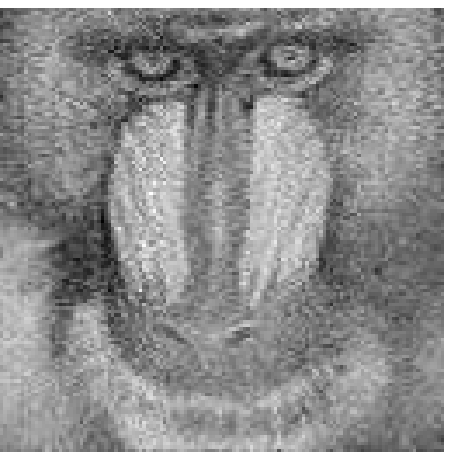}
\caption{JPEG Quality 70 ($p=22.37$)} \label{e13}
\end{minipage}
\end{tabular}
\end{center}
\end{figure}

\begin{figure}
\begin{center}
\begin{tabular}{cc}
\begin{minipage}{1.2in}
\centering
\includegraphics[width=1.0in,height=1.0in]{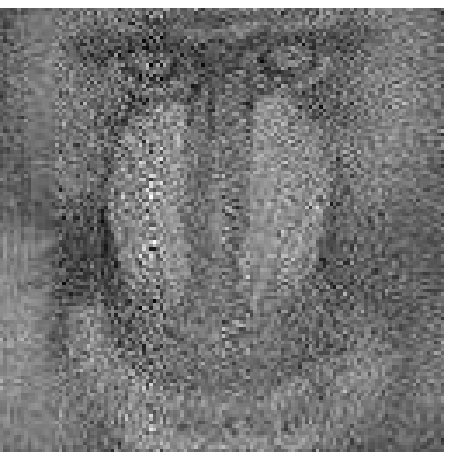}
\caption{JPEG Quality 50 ($p=16.75$)} \label{e14}
\end{minipage}
&
\begin{minipage}{1.2in}
\centering
\includegraphics[width=1.0in,height=1.0in]{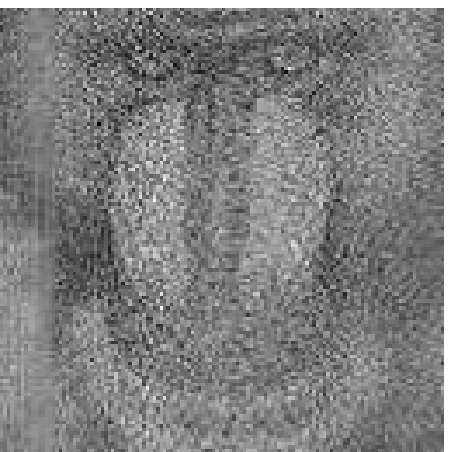}
\caption{JPEG Quality 30 ($p=17.42$)} \label{e15}
\end{minipage}
\end{tabular}
\end{center}
\end{figure}

JPEG is one of the image encoding schemes that is currently popular
[Wallace (1991)]. It is a lossy compression technique that uses
discrete cosine transform or the wavelet transform (2000 standard)
as the work horse that ensures a compressed image [Christopoulos et
al. (2000)]. We show that our watermarking scheme is robust against
JPEG compression to a great extent. In figures
(\ref{e12}-\ref{e15}), we show the watermark extracted from JPEG
compressed Lena image with various quality factors. Briefly, JPEG
quality factor is an indication of the distortion, such that 100\%
quality factor corresponds to least distortion. It is worth
mentioning here that, in our extraction algorithm as the JPEG
quality increases the corresponding PSNR values also increase
leading to a better extracted image. This fact is depicted in Fig.
(\ref{e11}). This, in turn, corresponds roughly to the amount of
information retained after wavelet or DCT decomposition. In Fig.
(\ref{e3}), we apply our watermark embedding algorithm with
$\lambda=0.2$ and we see that even at 50\% and 30\% JPEG quality
factors, the extracted watermark carries a reasonable resemblance of
the original watermark image. The PSNR values of $\mathbf{\Delta_w}
= |\mathbf{W - \widetilde{W}}|$ are provided as caption and as
expected they decrease with the decrease in quality factor. Hence,
the SVD based algorithm represented by equations
(\ref{eq2}-\ref{img_zc}) is robust against JPEG compression.

\section{Simulations for color image}

\begin{figure}
\begin{center}
\begin{tabular}{cc}
\begin{minipage}{1.2in}
\centering
\includegraphics[width=1.0in,height=1.0in]{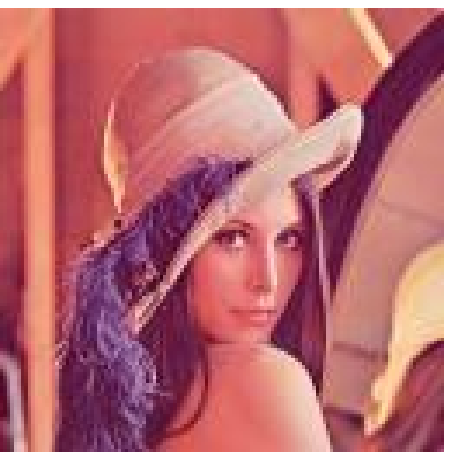}
\caption{Watermarked image} \label{e18}
\end{minipage}
&
\begin{minipage}{1.2in}
\centering
\includegraphics[width=1.0in,height=1.0in]{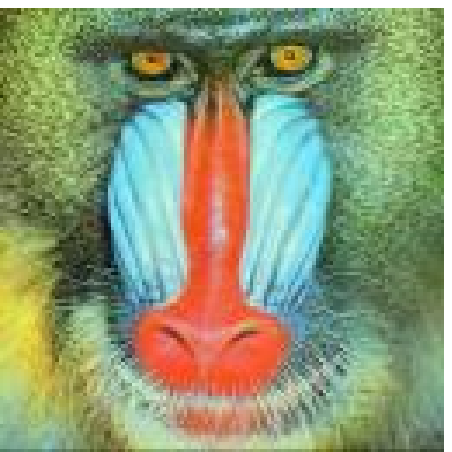}
\caption{Recovered watermark image} \label{e19}
\end{minipage}
\end{tabular}
\end{center}
\end{figure}

In Fig.(\ref{e18}), we show the simulations for the color images of
Lena and the baboon. Clearly, at $\lambda=0.02$, the watermarked
image does not show any sign of the underlying watermark and it is
invisible. The watermarking and extraction algorithm presented in
sections (2.1, 2.2) can be applied to color images as well. Since
most of the robustness can be generalized in a straightforward way
from those for the grey scale images, we do not show the results of
those tests here. However, we stress that the results of the
robustness test are similar to the ones for the grey scale images
presented above.

\begin{figure}
\begin{center}
\begin{tabular}{cc}
\begin{minipage}{1.2in}
\centering
\includegraphics[width=1.0in,height=1.0in]{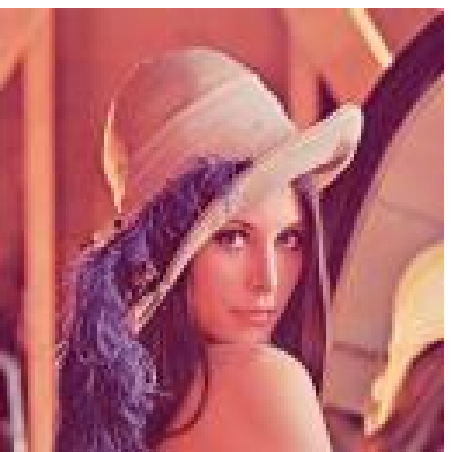}
\caption{original image} \label{e16}
\end{minipage}
&
\begin{minipage}{1.2in}
\centering
\includegraphics[width=1.0in,height=1.0in]{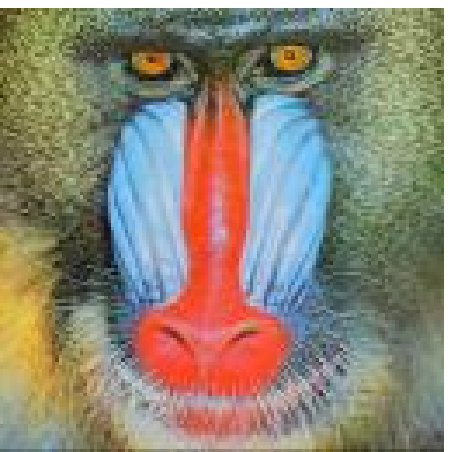}
\caption{Watermark image} \label{e17}
\end{minipage}
\end{tabular}
\end{center}
\end{figure}

\section{Conclusions}

We have presented an algorithm for digital watermarking using
singular value decomposition in the domain of the singular vectors
of the image matrices. We have implemented our method for gray and
color images and shown that the technique is robust against noise,
cropping and JPEG compression. The advantage of our algorithm, as
opposed to other SVD based techniques, lies in the fact that most of
the previous methods rely on the singular values of the image
matrix, thereby the information content of an $n \times n$ image was
translated into just $n$ values. In our approach, we use the scaled
singular vectors to encode the watermark. The technique also
inherently contains a level of security against the hacking of the
watermark.

The algorithm presented in the manuscript is not closed. In fact
there can be certain modification to the algorithm such that various
other aspects can be included. For example, one can decrease the
computational time of embedding and extracting the watermarks by a
segmentation of the image. Then the algorithm can be applied to
these various segments individually. Apart from this the embedded
image can be encrypted thereby enhancing the security of the image
content. Here, not only image data, but also other data can be
embedded in a cryptic form. Finally, it must be mentioned that in
the present algorithm the scaling factor $\lambda$ is not wholly
integrated into the algorithm, meaning if one does not possess the
value $\lambda$ then, extraction of the image becomes difficult. To
do away with this one can include the scaling factor inside the
image matrix itself. Currently the above mentioned features are
under investigation and we hope to report them soon.

\section{Acknowledgement}

One of the authors (RA) thanks Physical Research Laboratory
for the internship during which this work was done.


\vskip0.5cm

\textbf{References}

\vskip0.5cm

Andrews, H.C., Patterson, C.L., 1976. Singular Value Decomposition
(SVD) image coding. IEEE Trans. on communications. 24, 425-432.\\

Bors, A.G., Pitas, I., 1996. Image watermarking using DCT domain
constraints. IEEE-ICIP. 3, 231-234.\\

Chandra, D.V.S., 2002. Digital Image Watermarking Using Singular
Value Decomposition. Proceedings of 45th IEEE Midwest Symposium on
Circuits and System. 3, 264-267.\\

Chang, C.C., Tsai, p., Lin, C.C., 2005. SVD-based digital image
watermarking scheme. Pattern Recognition Letters. 26, 1577-1586.\\

Christopoulos, C., Askelöf, J., Larsson, M., 2000. Efficient methods
for encoding regions of interest  in the upcoming JPEG2000 still
image coding standard. IEEE Signal Processing Letters. 7, 247-249.\\

Craver, S., Memon, N., Yeo, B.L., Yeung, M.M., 1998. Resolving
rightful ownerships with invisible watermarking techniques:
Limitations, attacks and implications. IEEE Journal on Selected
areas in Communications. 16, 573-586.\\

Davidson, J.L., and Allen, C., 1998. Steganography using the Minimax
Eigenvalue Decomposition. Mathematical Methods of Data Coding,
Compression and Encryption, M. Schmalz, Ed., Proceedings of SPIE.
3456, 13-24. \\

Dickinson, T.B., 1997. Adaptive watermarking in the DCT domain. IEEE
ICASSP. 4, 2985-2988.\\

Ganic, E., Zubair, N., Eskicioglu, A.M., 2003. An Optimal
Watermarking Scheme Based on Singular Value Decomposition.
Proceedings of the IASTED International Conference on Communication,
Network, and Information Security. 85-90.\\

Golub, G.H.,Reinsch, C., 1970. Singular value decomposition and
least squares solutions. Numer. Math. 14, 403-420.

Gorodetski, V.I., Popyack, L.J., Samoilov, V., Skormin, V.A., 2001.
SVD-based Approach to Transparent Embedding Data into Digital Image.
International workshop on Mathematical methods, Models and
Architectures for Computer Network security. 2052, 263-274.\\

Katzenbeisser, S., Petitcolas, F.A.P., 2000. Information Hiding
Techniques for Steganography and digital Watermarking. Artech House. 121-148\\

Kundur, D., Hatzinakos, D., 1997. A robust digital image
watermarking scheme using the wavelet based fusion. IEEE-ICIP. 1,
544-547.\\

Kundur, D., Hatzinakos. D., 1998. Digital watermarking using
multiresolution wavelet decomposition. IEEE ICASSP. 5, 2659-2662.\\

Lacy, J.B., Quackenbusch, S.Q., Reibman, A.R., Shur, D.H., Snyder,
J.H., 1998. On combining watermarking with perceptual coding. IEEE
ICASSP. 6, 3725-3728.\\

Langelaar, G., Setyawan, I., Lagendijk, R.L., 2000. Watermarking
Digital Image and Video Data. In IEEE Signal Processing Magazine.
17, 20-43.\\

Leon, S.J., 1994. Linear Algebra with Applications, 4th edition,
Macmillan, New York.\\

Liu, R., Tan, T., 2002. A SVD-Based watermarking Scheme for
Protecting Rightful Ownership. IEEE Transactions on Multimedia.
4, 121-128.\\

Ohbuchi, R., Mukaiyama, A., Takahashi, S., 2002. A frequency-domain
approach to watermarking 3D shapes. Computer
Graphics Forum. 21, 373-382.\\

Petitcolas, F.B., Anderson, R.J., Kuhn, M., 1999. Information hiding
-A survey. Proc. of the IEEE. 87, 1062-1078.\\

Piva, A., Barni, M., Bartolini, F., Capellini, V., 1997. DCT-based
watermark recovering without resorting to the uncorrupted original
image. IEEE-ICIP. 1, 520-523.\\

Piva, A., Barni, M., Bartolini, F., 1998. Copyright protection of
digital images by means of frequency domain watermarking.
Mathematics of Data /Image coding, compression and Encryption, SPIE.
3456, 25-35.\\

Prandoni, P., Vetterli, M., 1998. Perceptually hidden data
transmission over audio signals. IEEE ICASSP. 6, 3665-3668.\\

Shieh, J.M., Lou, D.C., Chang, M.C., 2005. A semi-blind digital
watermarking scheme based on singular value decomposition. Computer
Standards and  Interfaces. Article in press.\\

Strang, G., 1993. The fundamental theorem of linear algebra,
American Mathematical Monthly. 100, 848-855.\\

Wallace, G.K., 1991. The JPEG still image compression standard.
Communications of the ACM. 34, 30-44.\\

Zeng, W., 1998. Digital watermarking and data hiding technologies
and applications. Proc. of ICISAS. 3, 223-229.\\

Zeng, W., Liu, B., 1999. A statistical watermark detection technique
without using original images for resolving rightful ownerships of
digital images. IEEE Trans. Image Processing. 8, 1534-1548.\\


\end{document}